\documentclass[pmlr,twocolumn,10pt]{jmlr} %

\usepackage{cite}
\usepackage{amsmath,amssymb,amsfonts}
\usepackage{textcomp}
\usepackage{xcolor}
\usepackage{bbm}
\usepackage{multirow}
\usepackage{soul}
\usepackage{booktabs}

\usepackage[switch]{lineno}
\jmlrvolume{259}
\jmlryear{2024}
\jmlrsubmitted{LEAVE UNSET}
\jmlrpublished{LEAVE UNSET}
\jmlrworkshop{Machine Learning for Health (ML4H) 2024} %

\title{Fundus Image-based Visual Acuity Assessment with PAC-Guarantees}

\begin{document}

\author{
\Name{Sooyong Jang}\Email{sooyong@seas.upenn.edu}\\
\Name{Kuk Jin Jang}\Email{jangkj@seas.upenn.edu}\\
\Name{Hyonyoung Choi}\Email{hyounchoi@seas.upenn.edu}\\
\addr{Computer and Information Science, University of Pennsylvania, USA}\\
\Name{Yong-Seop Han}\Email{medcabin@hanmail.net}\\
\addr{Department of Ophthalmology, Gyeongsang National University College of Medicine, Republic of Korea}\\
\Name{Seongjin Lee}\Email{insight@gnu.ac.kr}\\
\Name{Jin-hyun Kim}\Email{jim.kim@gnu.ac.kr}\\
\addr{Department of AI Convergence Engineering, Gyeongsang National University, Republic of Korea}\\
\Name{Insup Lee}\Email{lee@seas.upenn.edu}\\
\addr{Computer and Information Science, University of Pennsylvania, USA}
}

\maketitle

\def\Ds{\mathcal{D}}
\def\Hs{\mathcal{H}}
\def\Os{\mathcal{O}}
\def\Ss{\mathcal{S}}
\def\Ts{\mathcal{T}}
\def\Us{\mathcal{U}}
\def\Xs{\mathcal{X}}

\def\Rbbm{\mathbbm{R}}
\def\Prob{\mathbbm{P}}

\def\dh{\hat{d}}
\def\fh{\hat{f}}
\def\gh{\hat{g}}
\def\pih{\hat{\pi}}
\def\tauh{\hat{\tau}}
\def\gammah{\hat{\gamma}}
\def\tauh{\hat{\tau}}

\def\ie{\emph{i.e.}}
\def\eg{\emph{e.g.}}

\providecommand\SJ[1]{{\color{brown}[SJ: {#1}]}}
\providecommand\KJ[1]{{\color{blue}[KJ: {#1}]}}
\begin{abstract}
Timely detection and treatment are essential for maintaining eye health.
Visual acuity (VA), which measures the clarity of vision at a distance, is a crucial metric for managing eye health.
Machine learning (ML) techniques have been introduced to assist in VA measurement, potentially alleviating clinicians' workloads.
However, the inherent uncertainties in ML models make relying solely on them for VA prediction less than ideal.
The VA prediction task involves multiple sources of uncertainty, requiring more robust approaches.
A promising method is to build prediction sets or intervals rather than point estimates, offering coverage guarantees through techniques like conformal prediction and Probably Approximately Correct (PAC) prediction sets. 
Despite the potential, to date, these approaches have not been applied to the VA prediction task.
To address this, we propose a method for deriving prediction intervals for estimating visual acuity from fundus images with a PAC guarantee.
Our experimental results demonstrate that the PAC guarantees are upheld, with performance comparable to or better than that of two prior works that do not provide such guarantees.
\end{abstract}


\paragraph*{Data and Code Availability}
The Visual Acuity dataset was obtained from a previous study \citep{kim2022deep}
which consists of fundus images and corresponding visual acuity labels. 
This dataset is not publicly available due to restrictions on sharing patient images, but it is available from the corresponding author upon reasonable request and permission of the Institutional Review Board.
Code is available at this code repository.
\footnote{https://github.com/precise-ai4oph/va\_pred\_pac}

\paragraph*{Institutional Review Board (IRB)}
The appropriate approval was acquired for the study by the IRB at the institution where the data was collected. 
The procedures used in this study followed the principles of the Declaration of Helsinki. The requirement for informed patient consent was waived by the IRB due to the retrospective nature of the study. Additional details will be provided in the camera-ready version.

\section{Introduction}
\label{sec:introduction}

In eye health management, early detection and timely treatment are crucial.
Deep learning-based models have the potential to support large-scale screening programs, helping to detect abnormalities, support clinicians, and enable earlier diagnosis for individuals.
However, the inherent uncertainty in machine learning predictions challenges the effectiveness of such models in safety-critical applications like large-scale screening and clinician decision-support systems.

A common approach to addressing uncertainty is to 
predict a set (or interval) of labels with a Probably Approximately Correct (PAC) guarantee \citep{valiant1984theory} or to employ Conformal Prediction \citep{vovk2005algorithmic}, both of which provide formal guarantees on the reliability of predictions.
Machine learning approaches for analyzing fundus images—a type of retinal imaging—have been limited in providing formal guarantees compared to the vast amount of prior work in that field, which lacks such assurances
\citep{ayhan2020expert,zhou2022automorph,rahaman2021uncertainty,ge2021evaluation,filos2019systematic,leibig2017leveraging,ayhan2022clinical,gulshan2016development}.
\begin{figure}[tb]
\floatconts
    {fig:gt_obtain}
    {\caption{Uncertainties in VA prediction: Although VA prediction is based on fundus images, the ground truth acuity measurements were not obtained from these images. The VA test (e.g., using a Snellen chart) measures visual acuity, and the VA regressor estimates value measured by a human. Additionally, the human measurement process itself introduces uncertainty.}}
    {
    \includegraphics[width=0.9\columnwidth]{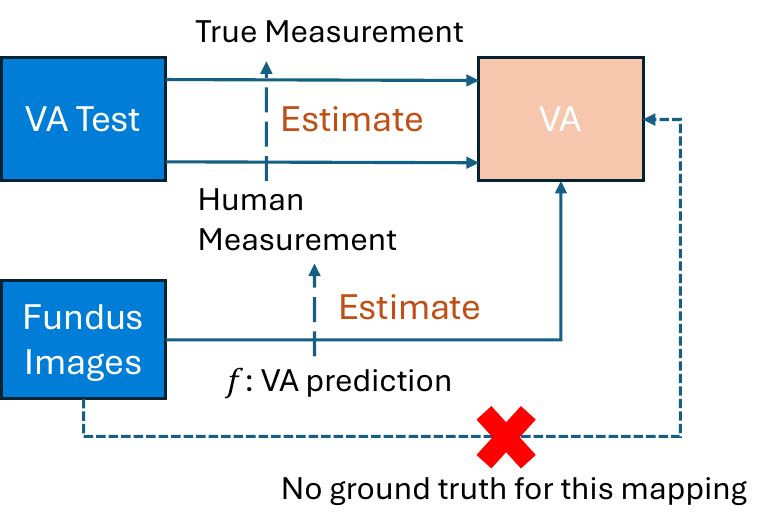} 
    }
\end{figure}

The situation is similar for visual acuity (VA) prediction, where the goal is to predict a subject's visual acuity based on fundus images.
The VA prediction task introduces uncertainties from multiple sources, as illustrated in Figure \ref{fig:gt_obtain}. 
First, the ground truth acuity value is obtained from a separate VA test conducted interactively with a human examiner and not derived directly from fundus images. 
Additionally, the VA test itself introduces variability and deviations from the true VA level \citep{siderov1999variability,vesely2012repeatability}. 
Given these various uncertainties, it is crucial to use prediction sets (or intervals) with guarantees for this task.
However, to the best of our knowledge, no previous studies have addressed this issue.

Two previous studies \citep{kim2022deep,paul2023accuracy} propose models to estimate visual acuity from fundus images.
Both studies evaluate their models on their respective datasets and demonstrate good performance. 
However, neither study provides any guarantees, which limits their clinical applications, such as in the aforementioned screening programs.
In this work, we address this limitation by constructing a prediction interval and providing a Probably Approximately Correct (PAC) guarantee in our model's predictions.
In detail, we train a regressor to estimate the visual acuity from fundus images, and derive a prediction interval $C(x_i)$ on example $(x_i, y_i)$ with a PAC guarantees on coverage, \ie{},
\begin{equation}
\label{eq:pac_guarantee}
    \Prob_{Z^n \sim \Ds^n} \left[ \Prob_{(x_i,y_i) \sim \Ds} \left[y_i \in C(x_i) \right] \ge 1 - \epsilon \right] \ge 1 - \delta,
\end{equation}
where $Z^n$ is the validation set, $\Ds$ is the data distribution, $\Ds^n=\Ds \times \dots \times \Ds$, and $\epsilon$ is an error bound and $\delta$ is a significance level.
Our experiments demonstrate that our approach achieves the target coverage while maintaining a sufficient average width of the prediction interval.
In summary, our contributions are as follows:
\begin{enumerate}
\item 
We propose a method for generating prediction intervals with example-dependent widths, ensuring a PAC guarantee by combining the well-established approach of training a model that outputs a Gaussian distribution with the PAC prediction interval technique.
\item 
We apply the aforementioned prediction interval method on a visual acuity dataset and empirically demonstrate that for a coverage bound $\epsilon=0.3$ and a significance level, $\delta=0.001~\%$, we achieve a coverage rate of 71.49 \% with average interval width of 3.04.
\end{enumerate}

The remainder of this paper is organized as follows:
We discuss related work in Section \ref{sec:relatedwork}, describe our method in Section \ref{sec:method}, and present our experimental results in Section \ref{sec:experiment}.
Finally, we conclude the paper in Section \ref{sec:conclusion}.

\begin{figure*}[tb]
\floatconts
    {fig:overall_process}
    {\caption{Overall process: First, we train a Gaussian Distribution output model with the training data. Next, we find $c$ for the coverage bound with the validation data. Lastly, we compute the prediction intervals using the test data.}}
    {
    \includegraphics[width=0.9\textwidth]{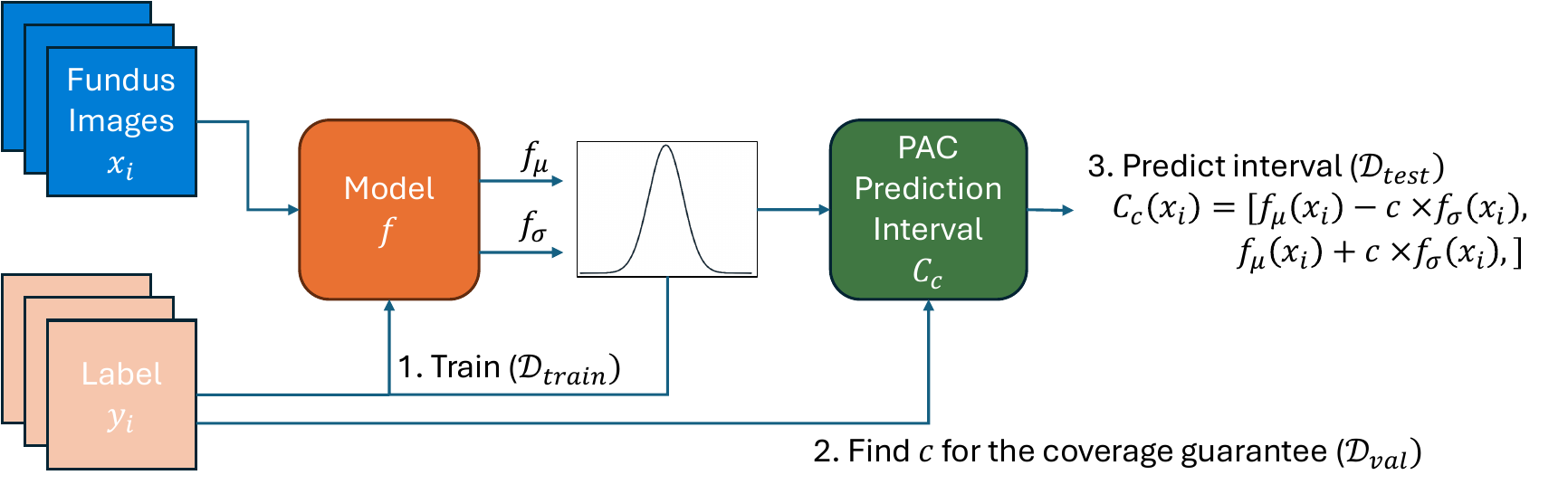} 
    }
\end{figure*}

\section{Related Work}
\label{sec:relatedwork}

In this section, we review the relevant literature. 
We begin by discussing two key studies focused on visual acuity prediction from fundus images.
Following this, we discuss prior work on prediction guarantees, with an emphasis on conformal prediction and PAC prediction sets.

\subsection{VA prediction from fundus images}
The Visual acuity (VA) prediction task is to estimate VA from fundus images.
VA can be represented in various ways, such as a decimal number (\eg{}, 0.5, 1.0), a fraction (\eg{}, 20/40, 20/20), or a letter score (\eg{}, 60, 35). 

The VA prediction has been explored in two recent studies \citep{kim2022deep, paul2023accuracy}.
In \citep{kim2022deep}, the problem was formulated as a classification task.
The original eleven visual acuity levels (ranging from 0.0 to 1.0 in increments of 0.1) were mapped into four levels based on an ophthalmologist's guidelines. They then employ a two-stage approach, with different classifiers at each stage. 
In contrast, \citet{paul2023accuracy} treat the problem as a regression task,
leveraging the ordinality of visual acuity labels to predict 
visual acuity letter scores using various models such as ResNet50, ConvNeXt, EfficientNetV2, and Swin Transformers. 
Both approaches demonstrate strong performance in their evaluations. However, neither study provides any guarantees on the reliability of their results.

\subsection{Conformal Prediction and PAC Guarantees}

One approach to providing guarantees on model predictions is to construct prediction intervals that, with high probability, include the true label.
This can be achieved using methods such as Conformal Prediction \citep{vovk2005algorithmic} and PAC Prediction sets \citep{vovk2012conditional, park2019pac}.
The two methods are similar, but offer different types of guarantees.
Conformal prediction generates a prediction interval that contains the ground truth for a test data point with high probability. In contrast, the PAC prediction set offers guarantees conditioned on the training data.

In VA prediction, an important consideration when providing guarantees on prediction intervals is the need for varying interval widths depending on the difficulty of each example. 
Ideally, easy examples (for the model) should have narrow prediction intervals, while more difficult examples should have wider intervals.
There has been extensive research in conformal prediction to address this need, using techniques such as scalar estimated uncertainty and locally adaptive conformal prediction \citep{papadopoulos2008normalized,romano2020classification,angelopoulos2021gentle,gibbs2021adaptive,seedat2023improving}. 
These approaches has been employed across various machine learning applications \citep{straitouri2023improving, ji2023incremental, park2022pac}.

When applying these techniques to the VA prediction task, a critical aspect is constructing intervals that have clinically useful widths. 
Specifically, the interval width must be narrow enough to be practical for VA prediction tasks. 
Although these techniques aim to minimize the interval width (or set size) while ensuring coverage, they do not address the practical requirements for interval width in clinical settings.

\section{Method} 
\label{sec:method}
{Our approach first trains a regression model for predicting VA from fundus images. The output is used to derive a prediction interval of the estimate with PAC guarantees. The overall process is illustrated in Figure \ref{fig:overall_process}.}

\subsection{Background - PAC Prediction Interval}
The PAC prediction interval (or set) constructs a prediction interval, $C(x)$ with a Probably Approximately Correct (PAC) guarantee, as represented in \equationref{eq:pac_guarantee}.
The PAC guarantee implies that the prediction error is small (``Approximately Correct", described by the inner 
probability with $\epsilon$), and holds with a high probability (``Probably", represented by the outer probability with $\delta$) as long as the test data follows the same distribution $\Ds$ as the training and calibration data.
In our context, ``Approximately" means that the prediction interval includes the true visual acuity with a coverage rate of at least $1-\epsilon$, while ``Probably" indicates that this coverage rate bound generally holds as long as all data—training, calibration, and test—are drawn from the same distribution, $\Ds$.

\subsection{Regression model learning}
\label{sec:method_learning}
Let $\Xs$ be the set of fundus images.
We train a model $f:\Xs \to \Rbbm^2$, to predict visual acuity for a given fundus image. 
We model the VA prediction with a Gaussian distribution,
where the mean represents the predicted visual acuity, and the standard deviation indicates the uncertainty of the prediction.
The two model output values correspond to the mean ($\mu \in \Rbbm$) and standard deviation ($\sigma \in \Rbbm_{> 0}$).
That is, for the $i^{th}$ fundus image $x_i$,  $f(x_i) = (f_{\mu}(x_i), f_{\sigma}(x_i)),$
where $f_{\mu}(x_i)$ and $f_{\sigma}(x_i)$ predict the mean and standard deviation, respectively.

To train this model, we use the Negative Log Likelihood (NLL) loss.
The standard deviation represents the prediction uncertainty and will be used to construct the prediction intervals, as illustrated in the following section.

\subsection{Prediction Interval with PAC guarantee}
{From the output, we build a prediction interval for VA,}
$C(x_i) = [C_l,C_u],$ 
{that should contain the true value with high probability, }
\ie{}, $y_i \in C(x_i)$.
{
Our goal is to derive the prediction interval that contains the ground truth with the minimum width, \ie{}, high coverage and narrow width, that satisfies the following PAC guarantee (\equationref{eq:pac_guarantee}).
}
With a constant $c$ which controls the prediction interval width satisfying \equationref{eq:pac_guarantee}, we derive
a prediction interval $C_c(x_i)$ for $(x_i,y_i)$, 
\begin{equation}
    C_c(x_i) = [C_{c,l}(x_i), C_{c,u}(x_i)],
\end{equation}
where $C_{c,l}(x_i) = f_{\mu}(x_i) - c \times f_{\sigma}(x_i), C_{c,u}(x_i)= f_{\mu}(x_i) + c \times f_{\sigma}(x_i)$, with a constant $c$ which 
{satisfies \equationref{eq:pac_guarantee}}.
Our interval uses the estimated standard deviation ($f_{\sigma}(\cdot)$) for each example, allowing each example to have a different prediction interval based on its estimated uncertainty (standard deviation).

We determine $c$  by solving the following optimization problem as described in \citep{park2019pac}:
\begin{align}
    c^* = \arg\min_{c} c \quad
    \text{subject to } \underline{c} \ge 1- \epsilon \nonumber,
\end{align}
where $[\underline{c}, \overline{c}]$ is the Clopper-Pearson interval for $W = \{\mathbbm{1}(y_i \in C_c(x_i))| (x_i,y_i) \in Z\}$ with the significance level $\delta$.
The prediction interval $C_{c^*}$ satisfies the PAC guarantee \citep{park2019pac, park2022pac}.

\section{Experiment}
\label{sec:experiment}
{
To evaluate our approach, we train models according to Section \ref{sec:method_learning} and compute prediction intervals satisfying the PAC guarantee. 
The experiments are repeated five times with different seeds, and we report the mean and standard deviations across the repetitions.
}

\subsection{Dataset}

{For training and evaluation, we use a fundus dataset obtained from existing work \citep{kim2022deep}.
The dataset consists of 54,781 fundus images labeled with visual acuity levels ranging from 0.0 to 1.0 in increments of 0.1, which were obtained through a visual acuity assessment.
}
{
The dataset uses categorized integer values from 0 to 10 (0, 1, 2, \dots, 10) to represent visual acuity levels ranging from 0.0 to 1.0, considering visual acuity prediction as a classification task.
We utilize these categorized labels for our regression task.
}
The details of the data distribution and related discussion are provided in \appendixref{apdx:dataset_imbalance}.
The dataset is randomly divided into training, validation, and test sets in a 6:2:2 ratio.
To evaluate the robustness, we create five different dataset splits of the dataset using different random seeds.

\subsection{Models}
We employ four different base models: Simple-CNN, ResNet18, ResNet50 \citep{he2016deep}, and EfficientNetV2-S \citep{tan2021efficientnetv2}.
{
All models have the final fully connected layer with two output nodes for the two Gaussian distribution parameters.
The Simple-CNN model that we implemented comprises two convolutional layers followed by three fully connected layers. 
All layers use ReLU activation functions, except for the final layer.
}
{
For the other models, we use pre-implemented versions available in PyTorch, modifying the final output layers to match the required dimensions.}
The ResNet and EfficientNet models are initialized with pre-trained weights provided by PyTorch, while the Simple-CNN is randomly initialized.
{
After the initialization, the models are trained using NLL loss as described in \sectionref{sec:method}.
}
\vspace{-1em}

\subsection{Results}
First, we present the performance of our regression model and with respect to the prediction intervals. In the subsequent section, we compare our results to prior work.

\begin{figure}[tb]
\floatconts
    {fig:res_mae}
    {\caption{MAE of models over 5 repetitions}}
    {
    \includegraphics[width=\columnwidth]{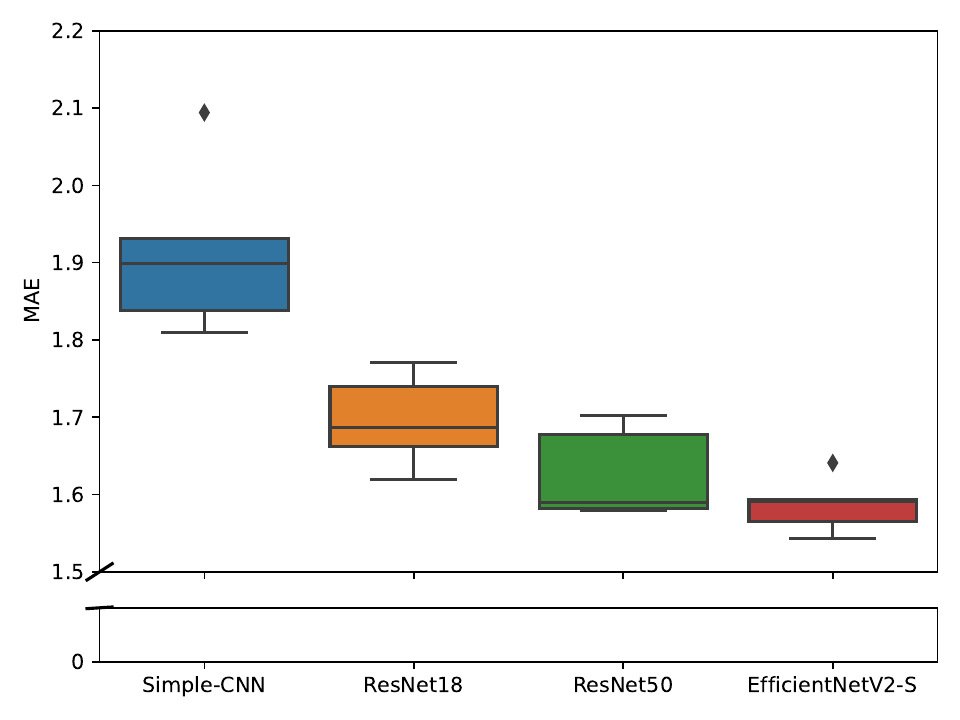} 
    }
\end{figure}

\subsubsection{Regression model}
The mean absolute error (MAE) of point predictions is displayed in \figureref{fig:res_mae}.
Based on the result, EfficientNetV2-S shows the best result with an average of 1.54, and the Simple-CNN shows the worst with an average of 1.88. 
While these MAEs might be acceptable considering the inherent inaccuracies in the (manual) assessment of visual acuity \citep{siderov1999variability, vesely2012repeatability},
it is important to note that the error can vary significantly for individual subjects because the models do not provide any guarantees.

\subsubsection{PAC prediction intervals}
\label{sec:pac_prediction_intervals}
We compute a prediction interval with PAC guarantee according to \sectionref{sec:method}.
We utilize a range of $\epsilon \in [0.2, 0.3, 0.4]$ 
for the coverage guarantee (\ie{} Coverage bound: $1-\epsilon$) 
 with a significance level $\delta$= 0.001~\%,
 and compute the coverage and the average interval width as shown in \figureref{fig:res_pac}.
Our method consistently satisfies the coverage bound for all $\epsilon$ values. 
All models show higher coverage than the given bound for all configurations. 

Based on the results, EfficientNetV2-S generates the narrowest prediction intervals, while Simple-CNN produces the widest. This variation may be partially due to differences in model complexity. More complex models, like EfficientNetV2-S, tend to learn intricate patterns in the data, allowing them to make more confident predictions with lower estimated standard deviation. As a result, these models produce narrower intervals. In contrast, simpler models, such as Simple-CNN, may have difficulty capturing such patterns, leading to greater uncertainty and wider prediction intervals.

Specifically, the average width for EfficientNetV2-S is around 3.0, when coverage rate bound is 70~\% ($\epsilon=0.3$), 
However, for practical usage with this 70~\% coverage, we require a slightly narrower width, around 2, which aligns with the variability in VA measurement by humans \citep{vesely2012repeatability}.  
\begin{figure*}[hp]
\floatconts
    {fig:res_pac}
    {\caption{PAC Guarantee Analysis Results
    {(significance level} $\delta$ = 0.001~\%). The left column displays the coverage, while the right column shows the average width. In the left column figures, the red dotted line represents the coverage bound. 
    For all $\epsilon$ and base models, our result show that the coverage bound is satisfied. 
    }}
{
    \subfigure [Coverage, $\epsilon=0.2$]{      
        \label{fig:res_pac_cov_0.2}
        \includegraphics[width=0.48\textwidth]{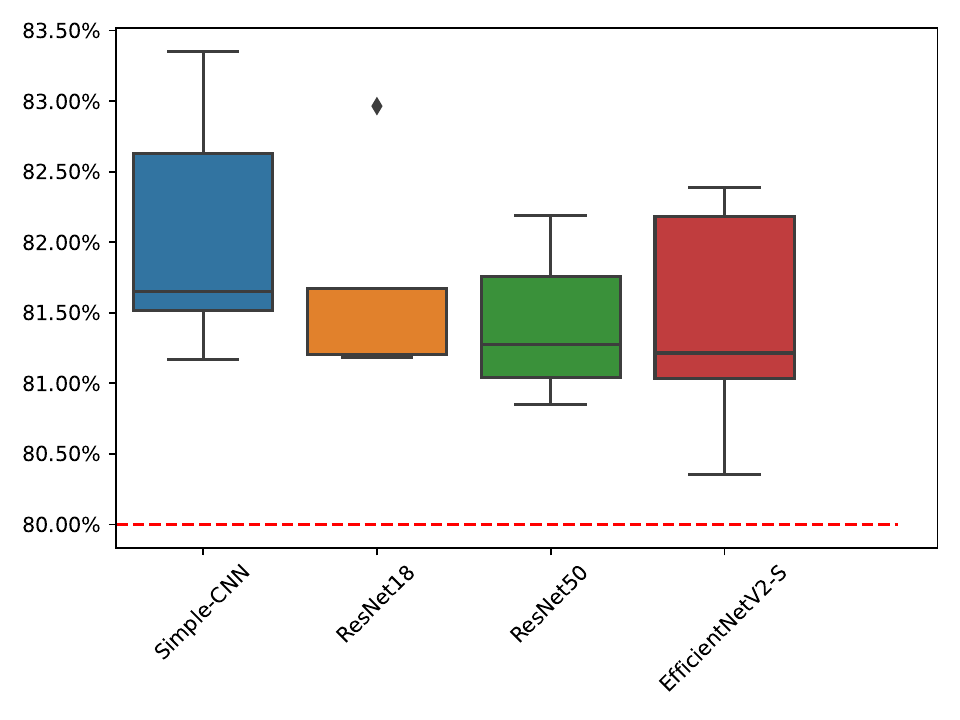}
    }
    \subfigure [Width, $\epsilon=0.2$] {
        \label{fig:res_pac_wid_0.2}
        \includegraphics[width=0.48\textwidth]{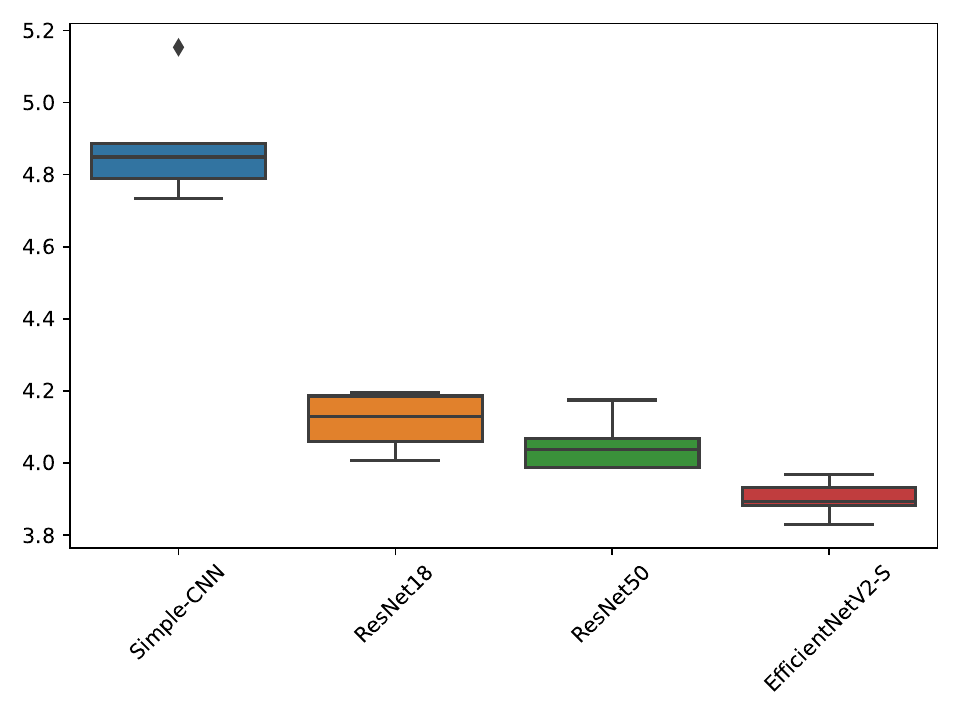}
    }
    \subfigure [Coverage, $\epsilon=0.3$] {
        \label{fig:res_pac_cov_0.3}
        \includegraphics[width=0.48\textwidth]{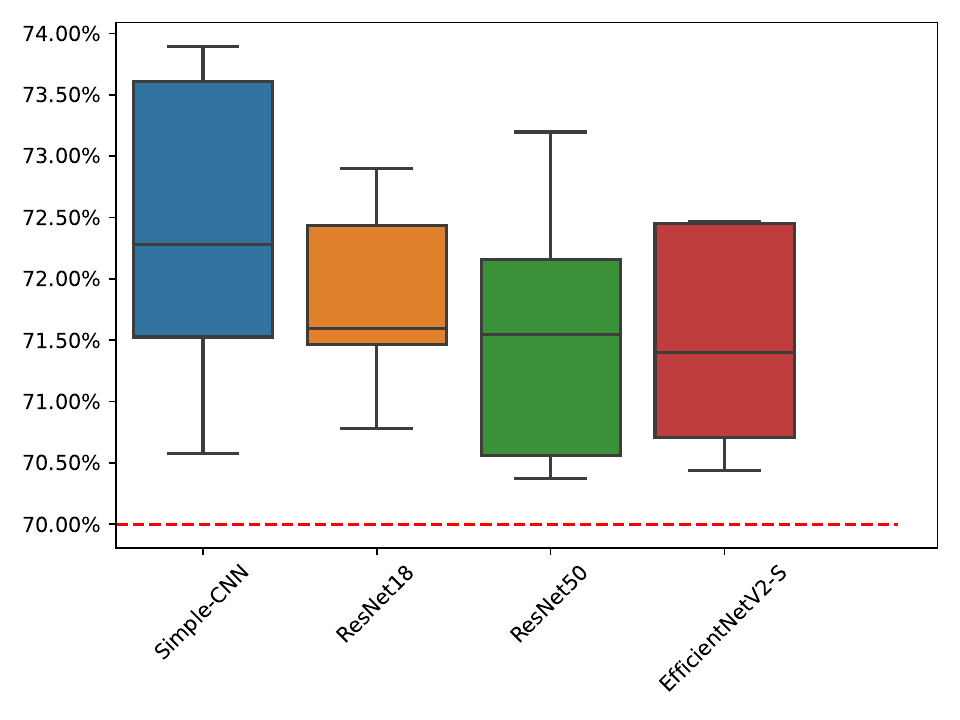}
    }
    \subfigure [Width, $\epsilon=0.3$] {
        \label{fig:res_pac_wid_0.3}
        \includegraphics[width=0.48\textwidth]{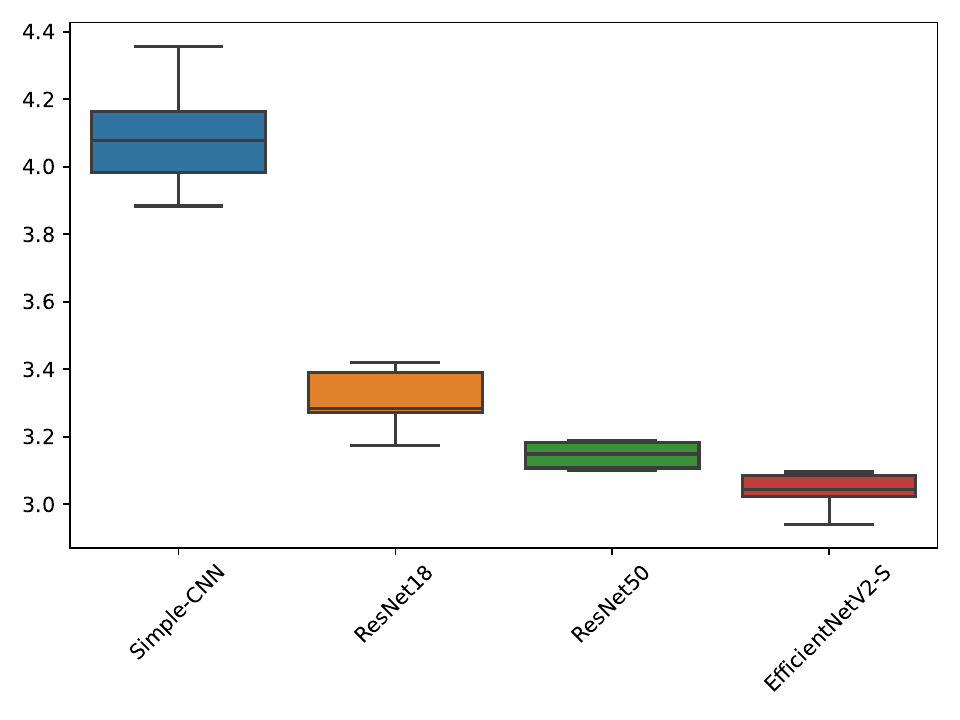}
    }
    \\
    \subfigure [Coverage, $\epsilon=0.4$] {
        \label{fig:res_pac_cov_0.4}
        \includegraphics[width=0.48\textwidth]{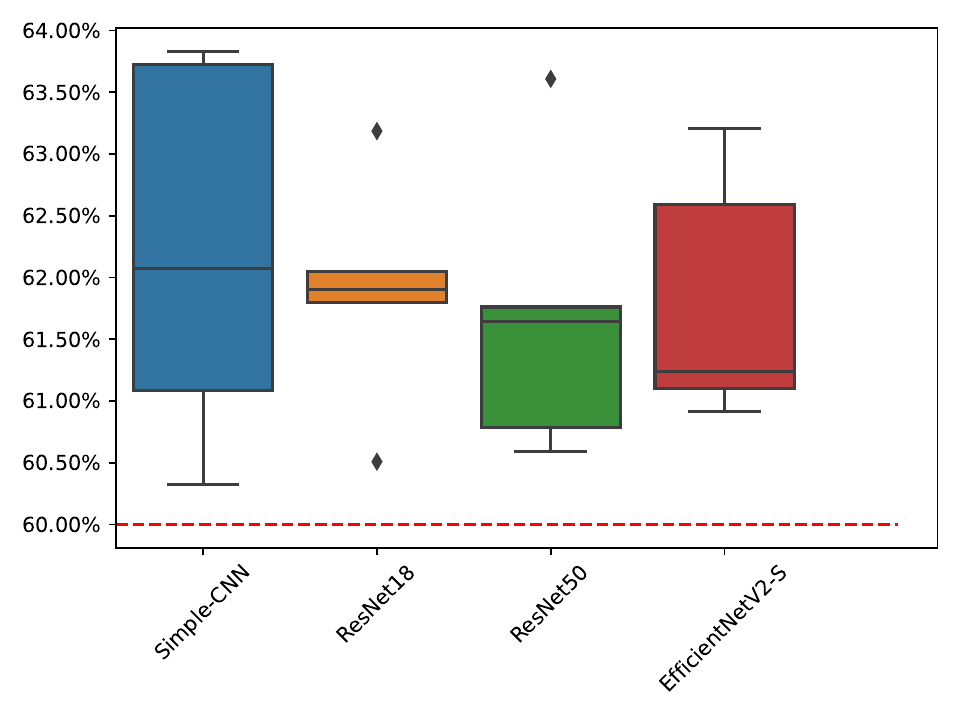}
    }
    \subfigure [Width, $\epsilon=0.4$] {
        \label{fig:res_pac_wid_0.4}
        \includegraphics[width=0.48\textwidth]{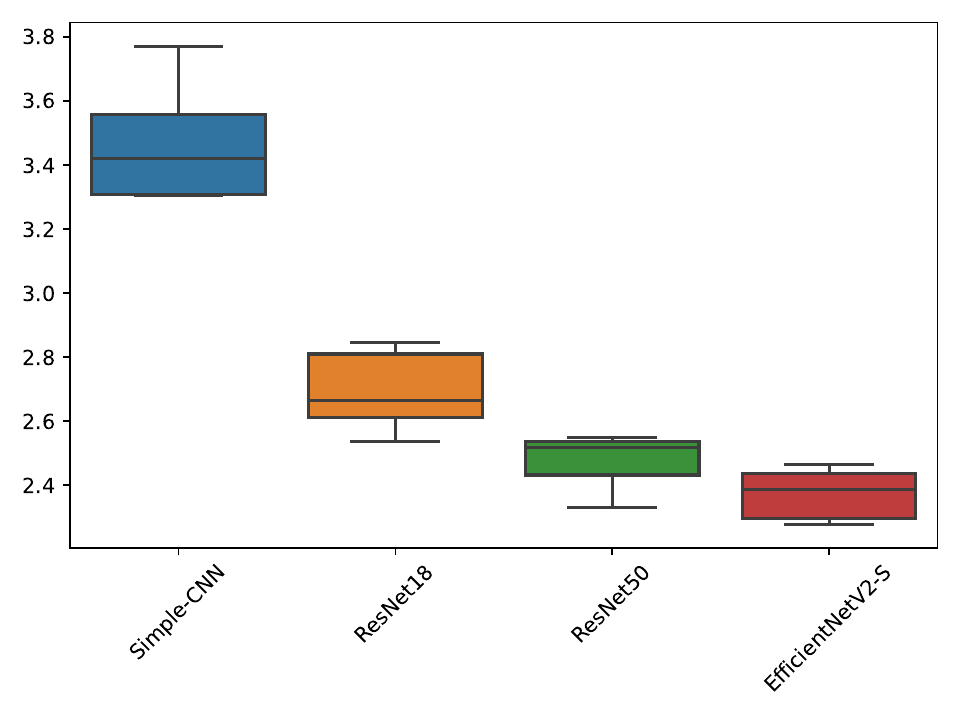}
    }
}
\end{figure*}

\subsection{Comparison to additional baselines}
We implement a Bayesian Neural Network (BNN) for ResNet18 and ResNet50 based on the code from the repository \citep{krishnan2022bayesiantorch} and vanilla conformal prediction as other baselines.
First, we compare the BNN performance for the point estimates with models from \sectionref{sec:method_learning}.
We compute the macro-averaged MAE to account for the dataset imbalance and compare the point estimate performance between our base model and the BNNs. 
This metric represents the average MAE for each ground truth value. 
As shown in \tableref{tab:cmp_bnn_point_estimate}, both BNN versions show similar point estimate performance compared to their deterministic counterparts. However, it is essential to note that the BNN approach does not provide a probabilistic coverage guarantee like our method.
\begin{table}[bt]
\centering
\floatconts {tab:cmp_bnn_point_estimate}
{\caption{Point Estimate Performance Comparison}}
{
\setlength{\tabcolsep}{3pt}
\begin{tabular}{ccc}
\toprule
Method & Base Model & Macro-Averaged MAE\\
\midrule
\multirow{4}{*}{Ours} & Simple-CNN & 2.51 $\pm$ 0.13\\
& ResNet18 & 2.10 $\pm$ 0.06\\
& ResNet50 & 2.14 $\pm$ 0.07\\
& EfficientNetV2-S & 2.12 $\pm$ 0.05\\
\midrule
\multirow{2}{*}{BNN} & ResNet18 & 2.17 $\pm$ 0.06\\
& ResNet50 & 2.29 $\pm$ 0.07\\
\bottomrule
\end{tabular}
}
\end{table}
To compare prediction interval performance, we perform Monte Carlo sampling to generate multiple predictions and compute the interval based on the 0.5 and 99.5 quantiles.
This enables us to then calculate coverage and interval widths.  
Since intervals from the BNN model do not ensure a coverage rate bound, the empirical coverage rate is not relevant to the expected quantile percentage (99\%). 
We also implement a vanilla conformal prediction (VCP) with the nonconformity score, $|\hat{y}_i - y_i|$. 
The results are presented in \tableref{tab:cmp_baselines}.
BNN interval widths are comparable to those of our models; however, they do not provide the guaranteed coverage.
VCP, designed to ensure coverage, does not always exceed the target rate of 70\% due to small fluctuations, which is expected \citep{angelopoulos2023conformal}. 
While the average interval width is comparable to our model's, this approach produces the same interval width for all examples. In contrast, our method provides adaptive widths based on the estimated standard deviation. 
\begin{table}[bt]
\centering
\floatconts {tab:cmp_baselines}
{\caption{Vanilla Conformal Prediction Performance}}
{
\setlength{\tabcolsep}{2.5pt}
\begin{tabular}{cccc}
\toprule
Method & Base Model & Coverage & Width\\
\midrule
\multirow{4}{*}{Ours} & Simple-CNN & 72.38 $\pm$ 1.25 & 4.09 $\pm$ 0.16\\
& ResNet18 & 71.83 $\pm$ 0.75 & 3.31 $\pm$ 0.09\\
& ResNet50 & 71.57 $\pm$ 1.04 & 3.15 $\pm$ 0.04\\
& EfficientNetV2-S & 71.49 $\pm$ 0.85 & 3.04 $\pm$ 0.06\\
\midrule
\multirow{2}{*}{BNN} & ResNet18 & 49.14 $\pm$ 2.69 & 3.03 $\pm$ 0.06 \\
& ResNet50 & 51.57 $\pm$ 4.12 & 3.11 $\pm$ 0.31\\
\midrule
\multirow{4}{*}{VCP}& SimpleCNN & 70.04 $\pm$ 0.71 & 3.94 $\pm$ 0.16\\
& ResNet18 & 69.72 $\pm$ 0.89
& 3.24 $\pm$ 0.09\\
& ResNet50 & 69.69 $\pm$ 1.13 & 3.07 $\pm$ 0.09\\
& EfficientNetV2-S & 69.43 $\pm$ 1.17 & 2.96 $\pm$ 0.07\\
\bottomrule
\end{tabular}
}
\end{table}

\vspace{-1.5em}

\subsection{Comparisons to previous studies}
We compare our approach with other methods for estimating visual acuity, as described in  \citep{kim2022deep, paul2023accuracy}.
These two approaches differ in their label scheme and the dataset, making a fair comparison challenging.
Nevertheless, we have made every effort to ensure the most accurate comparison possible.
\subsubsection{Comparison with \citep{kim2022deep}}
There are four main differences in the settings, a) Class scheme (4-level vs. 11-level), b) Dataset split (balanced test-set vs. imbalanced test-set), c) Prediction Type (Point vs. Interval), and d) Repetition (single experiments vs. repeated experiments).
{Considering the differences, we first map the visual acuity levels into 4 categories, as described by the authors in their paper (see Table \ref{tab:gnu_va_mapping}).
Next, we compute the macro-average accuracy (MA-ACC) of our prediction intervals on our imbalanced test set and compare it with the accuracy of their point predictions on the balanced test set.
For our MA-ACC calculation, we consider the interval correct when it contains a true label.}
In addition, we set $\epsilon=0.2$ for our intervals as per their final accuracy (82.4~\%). 
\begin{table}[bt]
\centering
\floatconts {tab:gnu_va_mapping}
{\caption{VA level mapping in \citep{kim2022deep}. Top row shows 11-level classes, bottom row shows corresponding 4-level classes.}}
{
\begin{tabular}{cc|cc|ccccc|ccc}
\toprule
11-level & 0 & 1 & 2 & 3 & 4 &5 &6 &7 & 8 & 9 & 10\\
\midrule
4-level & 0 & \multicolumn{2}{|c|}{1} & \multicolumn{5}{|c|}{2} & \multicolumn{3}{|c}{3}\\
\bottomrule
\end{tabular}
}
\end{table}
\begin{table}
\end{table}

{
The result is shown in Table \ref{tab:comp_gnu}. 
Our MA-ACC of prediction intervals is comparable with \citep{kim2022deep} with a width of around 4.37. 
In addition to comparable performance, we have the advantage of providing a guarantee on the coverage.
However, we note that a fair comparison is challenging due to differences in settings.
}

{
It should be noted that our MA-ACC in the table may be below 80~\% (1 - $\epsilon$), because the guarantee is not targeted for this metric. 
This metric is computed by macro-averaging classwise accuracy for the comparison to the \citep{kim2022deep}, while the prediction interval is computed to satisfy the coverage bound on the whole dataset.
Because of this difference, some of the computed MA-ACC in the table can be lower than the guaranteed coverage.
However, an appropriate interval could be computed if targeting the same metric.
}

\begin{table}[bt]
\centering
\floatconts {tab:comp_gnu}
{\caption{Accuracy comparison with \citep{kim2022deep}.}}
{
\begin{tabular}{ccc}
\toprule
    \multirow{2}{*}{Model} & \multicolumn{2}{c}{Metric}\\
     & MA-ACC & Width\\
\midrule
    Simple-CNN & 73.09 $\pm$ 5.63 & 6.05 $\pm$ 0.17\\
    ResNet18 & 79.27 $\pm$ 1.31 & 4.72 $\pm$ 0.07\\
    ResNet50 & 80.58 $\pm$ 2.66 & 4.54 $\pm$ 0.15\\
    EfficientNetV2-S & 80.77 $\pm$ 2.03 & 4.37 $\pm$ 0.15\\
\midrule
    \citet{kim2022deep} & 82.4~& - \\
\bottomrule
\end{tabular}
}
\end{table}

\subsubsection{Comparison with \citep{paul2023accuracy}}
We note two major differences from our settings in terms of a) Dataset and b) Visual Acuity Score System (Letter score vs. Fraction of Snellen chart).
{
Although the datasets are different, we compare the two results based on the error distributions reported in their paper. 
}
{
In \citep{paul2023accuracy}, they present the distribution of errors by computing the percentage of test examples that fall into three error ranges in the VA letter score:
between 0 and 5 ([0, 5]), between 6 and 10 ([6, 10]), and greater than 10 ([11,]).
We also compute the errors of our predictions using the same VA letter score and compare the error distributions across these ranges.
}
{
The difference in visual acuity scores was addressed using the conversion formula.\footnote{The conversion formula between the letter score ($L$) and the fraction ($F$) is: $L = 85 + 50 \times \log_{10}{F}.$}
}
\begin{table}[bt]
\setlength{\tabcolsep}{2pt}
\floatconts {tab:comp_jhu}
{\caption{Error distribution comparison with \citep{paul2023accuracy}. Each number shows the percentage of data in the error range.}}
{
\resizebox{\columnwidth}{!}{
\begin{tabular}{cccc}
\toprule
    \multirow{2}{*}{Model} & \multicolumn{3}{c}{Error Range}\\
     & [0,5] & [6,10] & [11,]\\
\midrule
    Simple-CNN & 58.09 $\pm$ 1.70 & 18.41 $\pm$ 1.66 & 23.50 $\pm$ 0.14\\
    ResNet18 & 60.97 $\pm$ 1.08 & 16.16 $\pm$ 0.59 & 22.87 $\pm$ 0.88\\
    ResNet50 & 60.56 $\pm$ 0.23 & 15.19 $\pm$ 0.54 & 24.25 $\pm$ 0.67\\
    EfficientNetV2-S & 61.55 $\pm$ 0.76 & 15.14 $\pm$ 0.27 & 23.31 $\pm$ 0.68\\
\midrule
    \citet{paul2023accuracy} & 33 & 28 & 39 \\
\bottomrule
\end{tabular}
}
}
\end{table}
{
As shown in Table~\ref{tab:comp_jhu}, our models have more examples in the small error range([0,5]) compared to the model in \citep{paul2023accuracy}
}
{Again,} based on these results, our model is at least comparable to or better than the model presented in \citep{paul2023accuracy} 
{while providing coverage guarantees.}

\subsection{
{
Discussion
}
}
\label{sec:future_work}
\begin{figure}[tb]
\floatconts
    {fig:res_mae_scale}
    {\caption{MAE and standard deviations: Absolute errors are binned by equal mass, and average error and standard deviations are computed for each bin, plotted with a 95~\% confidence interval.}}
    {
    \includegraphics[width=\columnwidth]{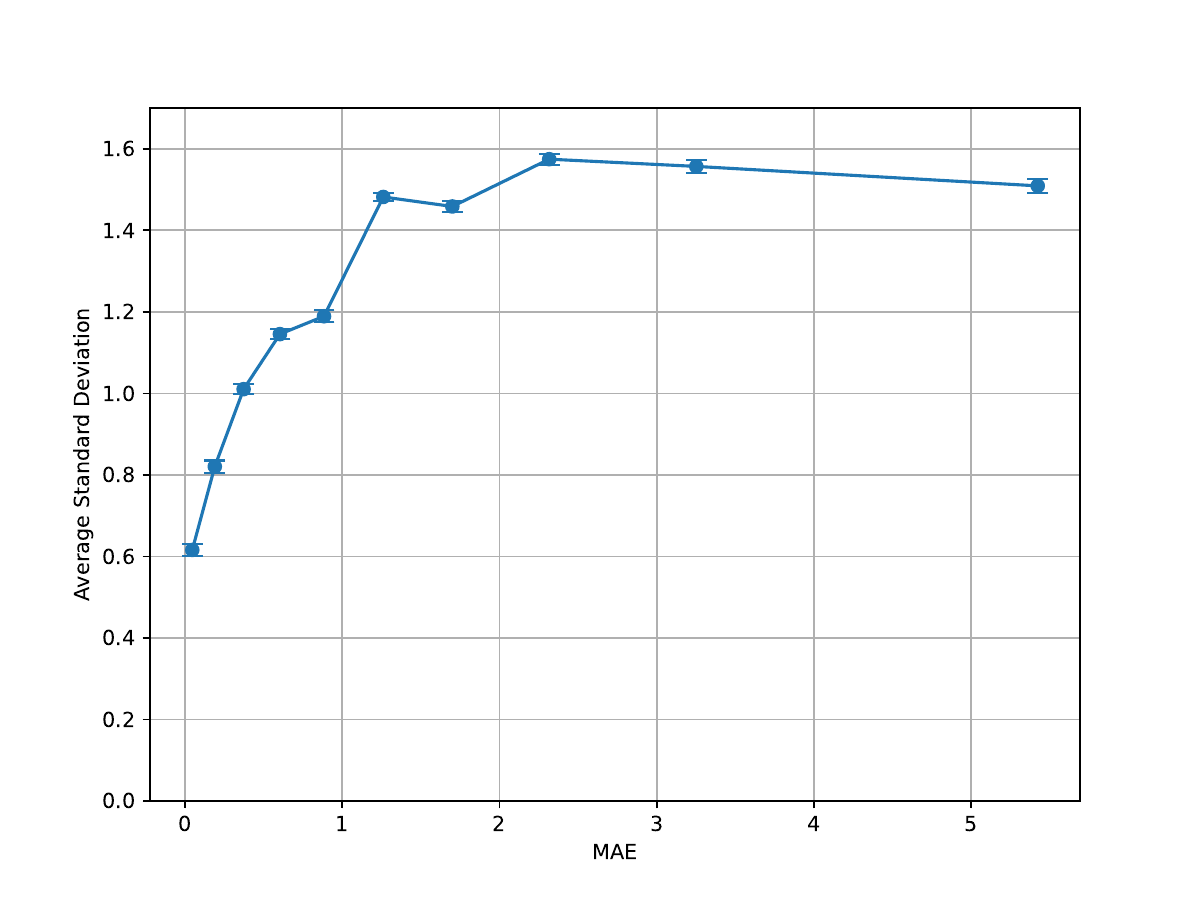} 
    }
\end{figure}

As shown in \sectionref{sec:pac_prediction_intervals}, our results indicate that the more complex model (EfficientNetV2-S) demonstrates better accuracy and produces narrower prediction intervals compared to less complex models. This suggests that more complex models may offer improved performance in terms of prediction interval width. We believe that models like EfficientNetV2-L and RETFound \citep{zhou2023foundation} are promising candidates for future work due to their higher complexity. In particular, as RETFound is a foundation model specifically designed for retinal images, it may be an especially suitable option.
Another point to
improve the prediction intervals in terms of the their width - narrower width, we examine the estimated standard deviation ($f_{\sigma}(x_i)$) of examples as our prediction interval width is derived from the standard deviations ($|C_c(x_i)| = 2 \times c \times f_{\sigma}(x_i)$). 
{
If the estimated standard deviation for each example are lower,
we can achieve narrower prediction intervals.
}

Figure \ref{fig:res_mae_scale} illustrates the relationship between the prediction error and estimated standard deviation for EfficientNetV2-S model using a random seed of 100. 
We compute both the prediction error ($|y_i - f_{\mu}(x_i)|$) and estimated standard deviation, then divide the prediction error space into five bins with equal mass, \emph{i.e.}, each bins contains the same number of examples. For each bin, we calculate the average absolute prediction error and the average standard deviation, which are then plotted in the figure.

The figure reveals a positive correlation between prediction error and standard deviation, indicating that examples with lower errors has the lower standard deviations. 
This observation suggests that for the subset of examples where the given model performs accurately, lower standard deviations can be achieved, resulting in narrower prediction intervals, because the prediction intervals are based on the predicted standard deviation.
It may be possible 
to identify such subsets of examples with low prediction errors, enabling us to provide narrower prediction intervals for the subset.
One significant challenge in applying Conformal Prediction and PAC based approaches in practice is the potential violation of their distributional assumptions. Specifically, the training (or calibration) data and test data may come from different distributions, a phenomenon known as dataset shift \citep{quinonero2022dataset}. To address this, several strategies have been proposed. Detection algorithms \citep{jang2022sequential, liu2020learning} can be employed to identify such shifts, enabling model retraining. 
Alternatively, adaptation algorithms \citep{sipac, parkpac} can adjust the model to account for the shifted distribution. These approaches can be applied to the visual acuity prediction problem. 
For instance, the adaptation algorithms in PAC prediction sets could be incorporated into our approach.
\textbf{Potential applications.}
Our approach can be extended to applications beyond visual acuity prediction. It can be applied to any regression task. Additionally, since the PAC interval can be used in classification tasks, this approach can be adapted for tasks such as diabetic retinopathy detection and glaucoma detection. In such classification tasks, the predictions will be sets of labels rather than intervals. For both regression and classification tasks, we can provide PAC guarantees for coverage.
\textbf{Ethical Considerations.}
Guidelines for this approach can be established, such as specifying the appropriate interval width based on the variability in traditional visual acuity measurements (around 2 within the 0-10 label range) \citep{vesely2012repeatability}.
If the predicted interval exceeds this width, its clinical usefulness may be diminished.
Furthermore, previous studies \citep{straitouri2023improving, babbar2022utility} have demonstrated that prediction intervals with probabilistic guarantees help clinicians make more accurate decisions. By engaging in meaningful discussions with clinicians about our approach, we believe that the effectiveness of our methods can be further maximized.
\textbf{Additional Analysis.}
We perform  qualitative and robustness analyses in \appendixref{apdx:additional_exp}. 
We visualize samples categorized by accurate vs. incorrect predictions and narrow vs. wide intervals, along with their activation maps. We also show that our algorithm remains robust under mild blurring conditions.

\section{Conclusion}
\label{sec:conclusion}

{
In this work, we developed a regression model that predicts visual acuity, modeled as a Gaussian distribution.
From the output, our approach derives prediction intervals with a PAC guarantee on coverage while varying the prediction interval widths.
Our empirical results demonstrate that the approach meets the probabilistic guarantee, with an average interval width of 3.04 for a coverage bound of $\epsilon=0.3$.
These outcomes are comparable to, or surpass, those of prior studies.
Importantly, our method provides a PAC guarantee on coverage, which is particularly advantageous for clinical applications - an aspect previously unaddressed.
This represents a significant step toward creating more trustworthy models for visual acuity prediction.
{
Future work could further enhance these models by: 1) employing more advanced base architectures, as our findings suggest that increased model complexity yields lower errors, and 2) leveraging the relationship between prediction error and estimated standard deviation, as discussed in Section \ref{sec:future_work}.
These future directions have the potential to lead to more accurate and clinically useful prediction intervals.
}
}

\acks{
This work was supported by the National Research Foundation of Korea(NRF) grant funded by the Korea government(MSIT) (No. NRF-2023R1A2C1006639).
Additionally, this research was supported in part by NIH 1R01EY037101 and ARO W911NF-20-1-0080.
}

\bibliography{references}

\clearpage

\appendix
\onecolumn
\renewcommand{\thetable}{\thesection.\arabic{table}} 
\renewcommand{\thefigure}{\thesection.\arabic{figure}} 

\section{Dataset Imbalance}
\label{apdx:dataset_imbalance}
The distribution of the data across each class is provided in Table \ref{tab:dataset}, and the dataset is imbalanced, with most of the data corresponding to good visual acuity (label = 9 or 10).
\begin{table*}[h]
\floatconts 
    {tab:dataset}
    {\caption{Dataset Overview: The dataset comprises 11 classes, with each class containing a varying number of data points.}}
    {\begin{tabular}{ccccccccccccc}
\toprule
    & 0 & 1 & 2 & 3 & 4 & 5 & 6 & 7 & 8 & 9 & 10 & Total\\
\midrule
    Count & 3274& 2164& 1647& 2091& 2005& 3240& 3803& 4261& 4370& 6358& 21568& 54781 \\
\bottomrule
\end{tabular}
}
\end{table*}
Although the dataset exhibits class imbalance, we did not apply oversampling or weighting techniques, as the PAC coverage still holds. The trade-off is that the intervals may be wider for minority classes due to the model's lower performance on these classes.
For comparison, we apply a resampling technique based on class frequency and evaluate the coverage rate and interval width of the EfficientNetV2-S model against those achieved with the vanilla technique.
As shown in Table \ref{tab:vanilla_resampling}, we do not observe significant improvements from resampling and therefore it is not included in the final version.
\begin{table}[h]
\centering
\floatconts {tab:vanilla_resampling}
{\caption{Comparison between the vanilla and resampling technique. $\epsilon=0.3, \delta=0.001 \%$}}
{
\begin{tabular}{ccc}
\toprule
Technique & Coverage Rate & Width\\
\midrule
Vanilla & 71.49 $\pm$ 0.85 &
3.04 $\pm$ 0.06\\
Resampling & 71.79 $\pm$ 1.35 & 3.21 $\pm$ 0.09\\
\bottomrule
\end{tabular}
}
\end{table}

\section{Additional Experiment Results}
\label{apdx:additional_exp}

\subsection{Qualitative Analysis}
\label{apdx:qualitative}
We conduct a qualitative analysis of our model’s results, specifically sampling images where the model’s prediction interval either contains the ground truth or does not, as well as images with both wide and narrow intervals. Additionally, we apply EigenGradCAM to analyze the model’s prediction behavior.
\\
\figureref{fig:res_qual_samples} presents sample images with different ground truth values, illustrating cases where the prediction intervals either cover or fail to cover the ground truth. 
The second figure displays sample images with varying ground truth values, comparing two scenarios: narrow prediction intervals ($\le$ 2.0) and wide intervals ($\ge$ 5.0).
Generally, fundus images with higher visual acuity (higher class) tend to be clearer. Additionally, we plot the activation maps from the neural network layers for deeper analysis.

\begin{figure}[h]
\floatconts
    {fig:res_qual_samples}
    {\caption{Image Samples for Different Cases.}}
{
    \subfigure[Covered vs. Not Covered]{
        \label{fig:res_cov_notcov_samples}
        \includegraphics[width=0.9\textwidth]{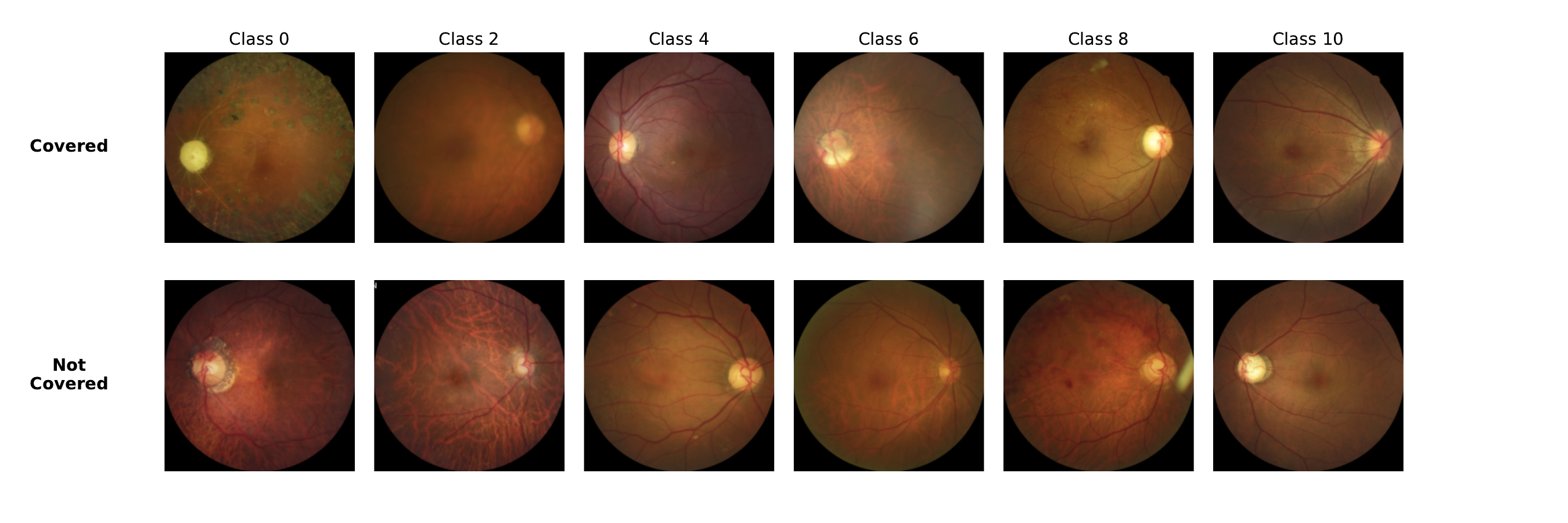}
    }
    \subfigure [Narrow vs. Wide]{
        \label{fig:narrow_wide_samples}
        \includegraphics[width=0.9\textwidth]{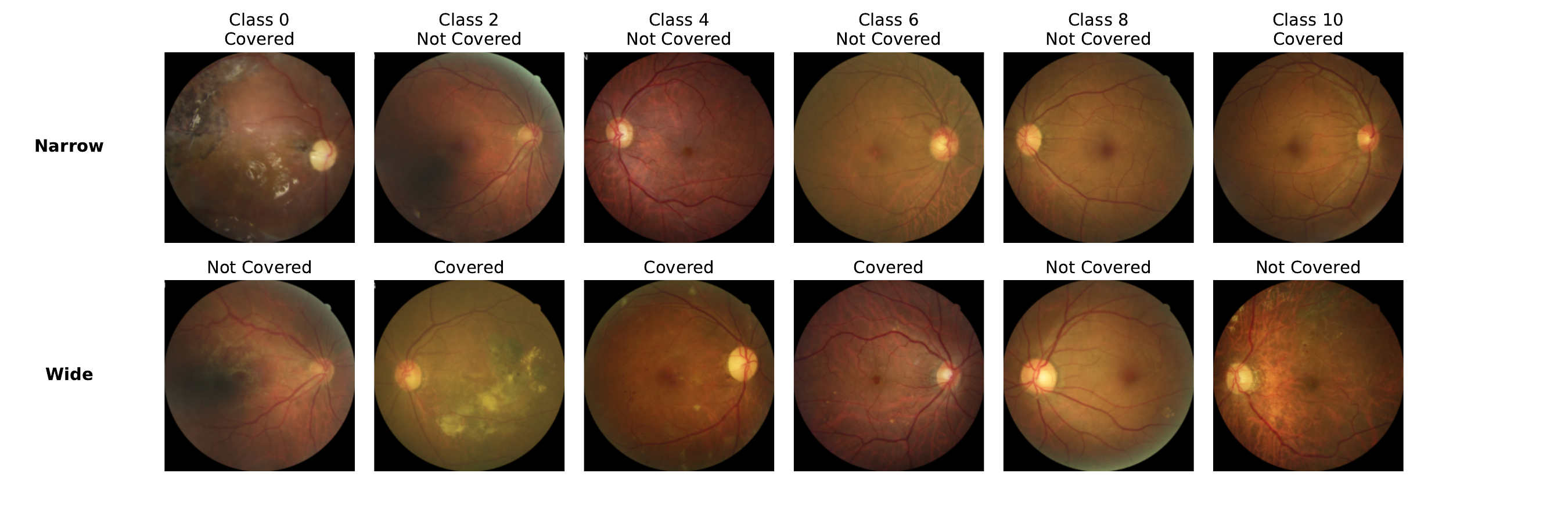}
    }
}
\end{figure}

\figureref{fig:res_gradcam_samples} shows activation maps generated by EigenGradCAM (using code from https://github.com/jacobgil/pytorch-grad-cam?tab=readme-ov-file) for the sample images drawn in the previous figures. These maps indicate that the model mainly focuses on the macula region. Furthermore, when the model evaluates a wider region, the prediction interval tends to include the ground truth, and the intervals are wider.

\begin{figure}[h]
\floatconts
    {fig:res_gradcam_samples}
    {\caption{Activation Maps from EigenGradCAM}}
    {\includegraphics[width=0.9\textwidth]{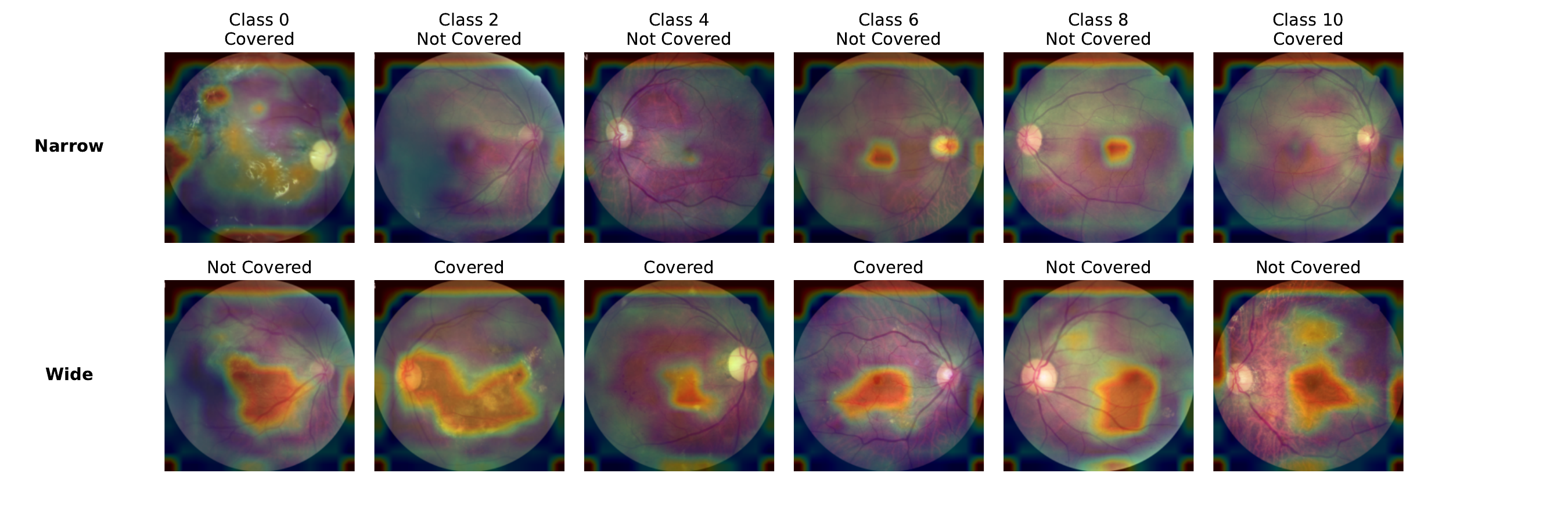}}
\end{figure}

\subsection{Robustness Analysis}
\label{apdx:robustness}
We conduct experiments with multiple repetitions using different dataset splits based on various seeds, reporting the average and standard deviation. 
We believe that this experimental setup demonstrates the robustness of our approach across different participants. Additionally, the PAC interval algorithm provides a conditional guarantee based on the training data (as indicated in \equationref{eq:pac_guarantee}, where the calibration set $Z^n$ is sampled), ensuring coverage regardless of the calibration set.
\\
In terms of image quality, our approach may yield wider intervals for low-quality images, which might not be clinically helpful. 
Achieving accurate predictions in such cases is quite challenging. 
However, we believe this issue can be addressed by utilizing automatic image quality detection algorithms \citep{raj2019fundus,coyner2019automated,karlsson2021automatic,abramovich2023fundusq,chuter2024deep}. These algorithms can alert users when a captured image is of poor quality and prompt them to retake it. Subsequently, our algorithm will operate on high-quality images.

\begin{figure}[h]
\floatconts
    {fig:res_robustness}
    {\caption{Robustness Analysis Under Varying Levels of Blur}}
{
    \subfigure[Robustness of Our Model]{
        \label{fig:res_robustness_level}
        \includegraphics[width=0.6\textwidth]{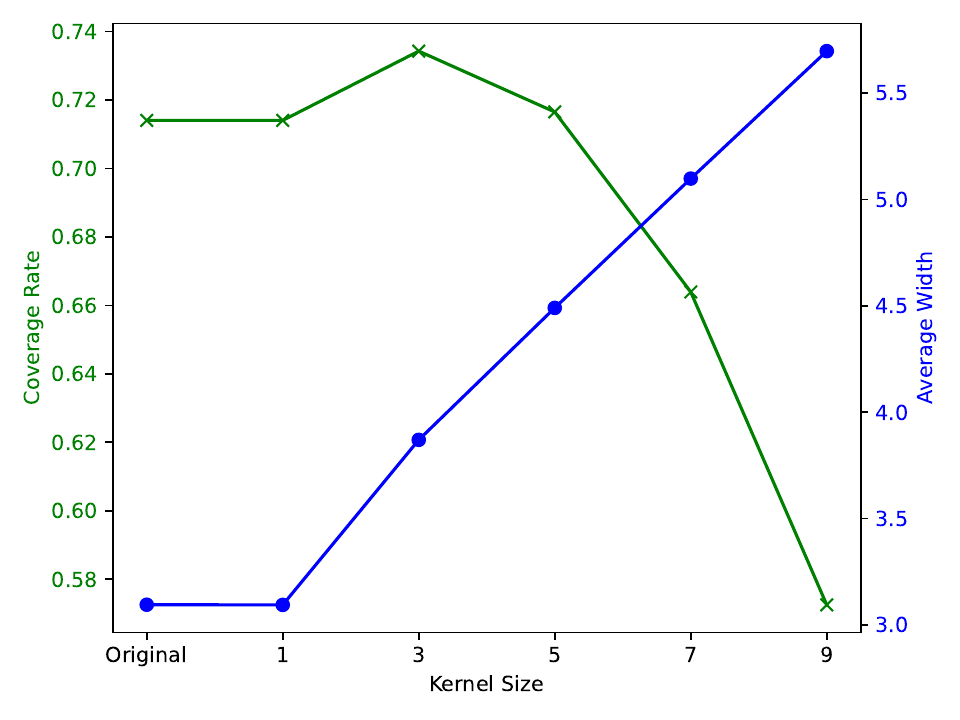}
    }
    \subfigure [Blurred mage Samples]{      
        \label{fig:blur_samples}
        \includegraphics[width=0.8\textwidth]{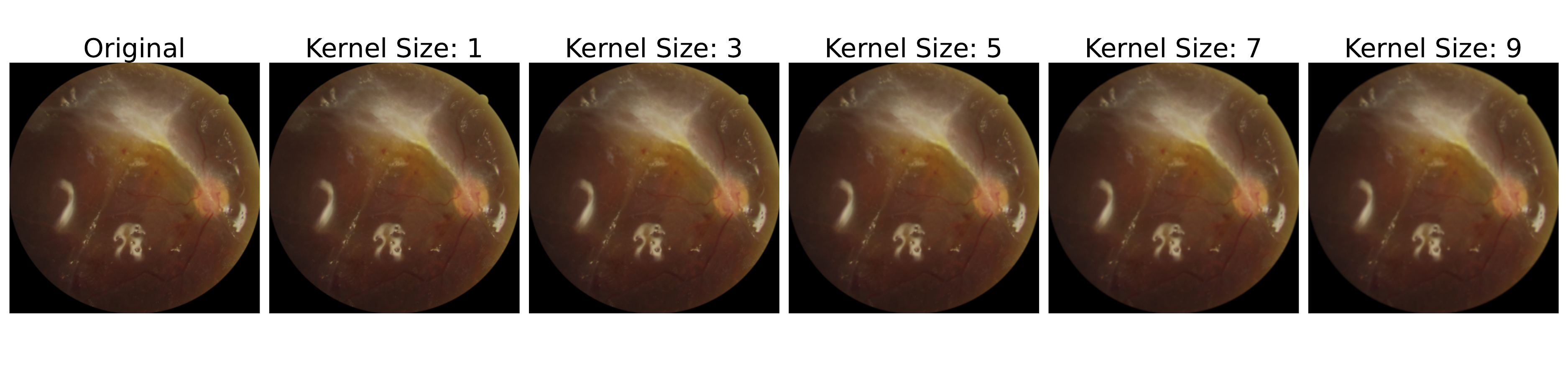}
    }
}
\end{figure}

Since severe image quality degradation can be easily detected by the image quality detection algorithms, we investigate how well our approach handles non-severe low-quality images. 
We simulate varying image qualities by applying Gaussian blur with different kernel sizes (keeping sigma fixed at 5), as illustrated in \figureref{fig:blur_samples}. 
The result with EfficientNetV2-S  (\figureref{fig:res_robustness_level}) indicates that up to a kernel size of 5, the coverage rate remains valid, though the average interval width increases. 
However, with more severe blurring, our model fails to maintain the coverage rate as the distributional assumptions are violated. 
These results evidence the robustness of our approach to image quality degradation. 

\end{document}